\documentclass[prb]{revtex4-1} 

\usepackage{amsmath}  % needed for \tfrac, \bmatrix, etc.
\usepackage{amsfonts} % needed for bold Greek, Fraktur, and blackboard bold
\usepackage{graphicx} % needed for figures
\usepackage{color}

\usepackage[latin1]{inputenc}

\begin{document}

\title{Hanbury Brown and Twiss effect demonstrated for sound waves from a waterfall; an experimental, numerical and analytical study}

\author{Arnt Inge Vistnes}
\email{a.i.vistnes@fys.uio.no} % optional
\altaffiliation[permanent address: ]{Kurlandslia 20, 1479 Kurland, Norway}
\affiliation{Department of Physics, University of Oslo, Sem S\ae lands vei 24, Oslo 3, Norway}
% Please provide a full mailing address here.

\author{Joakim Bergli}
\email{joakim.bergli@fys.uio.no}
\affiliation{Department of Physics, University of Oslo, Sem S\ae lands vei 24, Oslo 3, Norway}

% See the REVTeX documentation for more examples of author and affiliation lists.

\date{\today}

\begin{abstract}

The Hanbury Brown and Twiss effect (HBT) is described by numerical and analytical modeling, as well as experimentally, using sound waves and easily available instrumentation. An interesting phenomenon that has often been considered too difficult to be included in standard physics studies at bachelor and master level, can now be introduced even for second year bachelor students and up. 

In the original Hanbury Brown and Twiss effect the angular size of the source (the star Sirius) was calculated by determining the distance between two detectors that lead to a drop in the cross-correlations in the signals from the detectors. We find that this principle works equally well by sound waves from a waterfall. This is remarkable, since we use a completely different kind of waves from the HBT case, the frequency of the waves differ by a factor $\sim 10^{12}$ and the wavelength as well as the angular extension of the source seen from the observer's position differ by a factor $\sim 10^{7}$.

The original HBT papers were based on measurements of \emph{intensity} fluctuations recorded by two detectors and correlations between these signals. The starting point for the theory that explained the effect was therefore intensity fluctuations per see, and the theory is not easy to understand, at least not for an undergraduate physics student. The quantum description that followed a few years later was even more difficult to understand.

Our starting point is descriptions of broadband waves at the amplitude level (not at intensity level) by numerical modeling. Important properties of broadband waves can easily be revealed and understood by numerical modeling, and time-resolved frequency analysis (TFA) based on Morlet wavelets turns out to be a very useful tool. In fact, we think it has been far too little attention to broadband waves in physics education hitherto, but the growth of use of numerical methods in basic physics courses opens up new possibilities.

It was impossible in 1956 to record electromagnetic waves like visible light from the star Sirius at amplitude level (electric field vs time), and it still is. The frequency is far too high. However, broadband sound waves in air have so low frequencies that it is very easy to make recordings at the amplitude level. From such measurements, we can calculate intensity variations numerically, allowing us to study the HBT effect both at the amplitude and intensity level. It is this close connection between a description at the amplitude and intensity level, and focus on broadband waves, that is the key for our work. 

Our paper is long since it is, in a way, three papers in one. However, it is possible to skip the theoretical treatment (late in the paper) initially, and still get a good idea of our model.

Computer programs as well as original sound files are available from the authors.
\end{abstract}

\maketitle % title page is now complete

\section{Introduction}

In the 1950s Hanbury Brown and Twiss (HBT) developed a method where the angular extension of objects on the sky could be determined from time variation in intensity measurements at two nearby radio telescopes~\cite{HBT1954}. When the telescopes were close to each other, a high correlation of the signals from the two telescopes was observed, but the correlation dropped considerably as the distance between the telescopes increased beyond a certain limit. The angular extension of the source could be determined from these data based on a classical statistical wave description. 

Soon after, the method was also implemented at optical wavelengths in the measurement of the angular diameter of the star Sirius~\cite{HBT},\cite{TLHB1957}. The fact that the HBT effect also arise at optical wavelengths where the quantum nature of the electromagnetic field should become important ``initially encountered considerable resistance in the scientific community'', as Klaus Hentschel expresses it in his new book~\cite{Hentschel}.  However, after Glauber worked out the quantum theory of optical coherence~\cite{Glauber}, our impression is that many physicists consider the HBT effect as a true quantum phenomenon which leads to ``photon bunching''. The quantum as well as the semi-classical explanation of the phenomena (see e.g. Ref~\onlinecite{MandelWolf}) are rather complex both at the conceptual and mathematical level. The phenomenon is considered not suitable for undergraduate physics curricula. The HBT method seems not to be much used in the original context any more but it is common in high energy physic~\cite{Baym}, providing motivation for introducing this phenomenon to physics.

Innovative numerical modeling has lately been introduced in physics education at the undergraduate level, and offers a very valuable tool in addition to traditional experiments and theory. Numerical modeling often makes it easier to handle complex physics than theory alone, making the problems more relevant and interesting for the students. In addition, it also opens up for new conceptual pathways.~\cite{Morten} 

In this paper we use numerical modeling of the HBT effect. We think that this can lead to a better understanding not only of the HBT effect, but of broadband waves in general. By ``broadband waves'' we here mean waves with a rather broad continuous frequency distribution. In undergraduate physics, waves are often treated synonymously with simple harmonic waves. This limits in a serious way the student's richness of thoughts about waves. 

The central point is to become aware of key differences between a harmonic wave and a broadband wave. A single source may generate broadband waves, but often, broadband waves are a sum of the radiation from many independent part-sources. One example is electromagnetic waves from a star, as was studied in the original HBT experiment. We have performed experiments on sound waves from a waterfall as a more accessible acoustic analogue. Numerically, we can generate a continuous broadband wave by an inverse Fourier transform of a proper frequency distribution. 

The HBT effect appears in situations where many broadband waves add to each other at the position of the detectors. The summation of waves is slightly different at detector A's position compared to at the detector B's position when the detectors are separated. The reason for this is that the origins of the different broadband part-waves are distributed over a finite, limited region in space.

At the University of Oslo numerical modeling of the HBT effect was used as a one week student project for second year bachelor students in a course in oscillations and waves in the spring 2014~\cite{Vistnes2130}. 
Numerical modeling of the HBT effect for light has also been utilized as a part of a lab for undergraduates at Brigham Young University in 2015~\cite{Kingsley}. The HBT effect has also been used as a space positioning method based on sound fields~\cite{Zou}.

\section{Broadband waves studied by time-frequency analysis (TFA)}\label{TFAexpl}

In this section, we first imagine to have one source of a broadband wave. The wave is recorded by a detector at a sampling frequency well beyond twice the maximum frequency in the wave. Thus, the recording is performed at the 'amplitude level' (e.g. sound-pressure vs time).

This may seem to be a strange choice, since the Hanbury Brown and Twiss experiment on Sirius was performed at the intensity level, since the time resolution of the detectors and the following instrumentation was far too low compared to the frequency of visible light (and still is). However, waves always add to each other at the amplitude level, and we will show that \emph{the HBT effect is fundamentally an amplitude level effect}, but since there is a close connection between amplitudes and intensities, it will also be possible to use it at the intensity level as it was done historically. We will in this paper do both. 

We model a broadband wave numerically by generating a Fourier spectrum where each of the frequency components has random amplitude (limited by a Gaussian shape) and random phase.  The center frequency is $f_0$ and the full width at 1/e of max was chosen to be $f_0$. The relative width is then similar to the frequency distribution for visible light from Sirius, and similar to the filtered sound wave signals used in our experiments. By an inverse Fourier transform we get the amplitude vs time signal representing the wave.

The resulting wave can be described by
\begin{equation}
W(t) = \sum_{k=1}^{N} {a_k e^{i \omega_k t}} + c.c.
\label{eq:numBasic}
\end{equation}
\noindent
where $\omega_k$ is angular frequencies, and the coefficients $a_k$ are complex numbers with absolute values chosen randomly between zero and a maximum value given by a gaussian frequency distribution
\[
a_k = e^{-\frac{(\omega_k-\omega_0)^2}{2 \omega_0^2}}.
\label{eq:numFreq}
\]

In the numerical description, the time is also discretized, but we leave out those details from our expressions to avoid more complex notations than necessary.

\begin{figure}[h] 
  \centering
  \includegraphics[width=6cm]{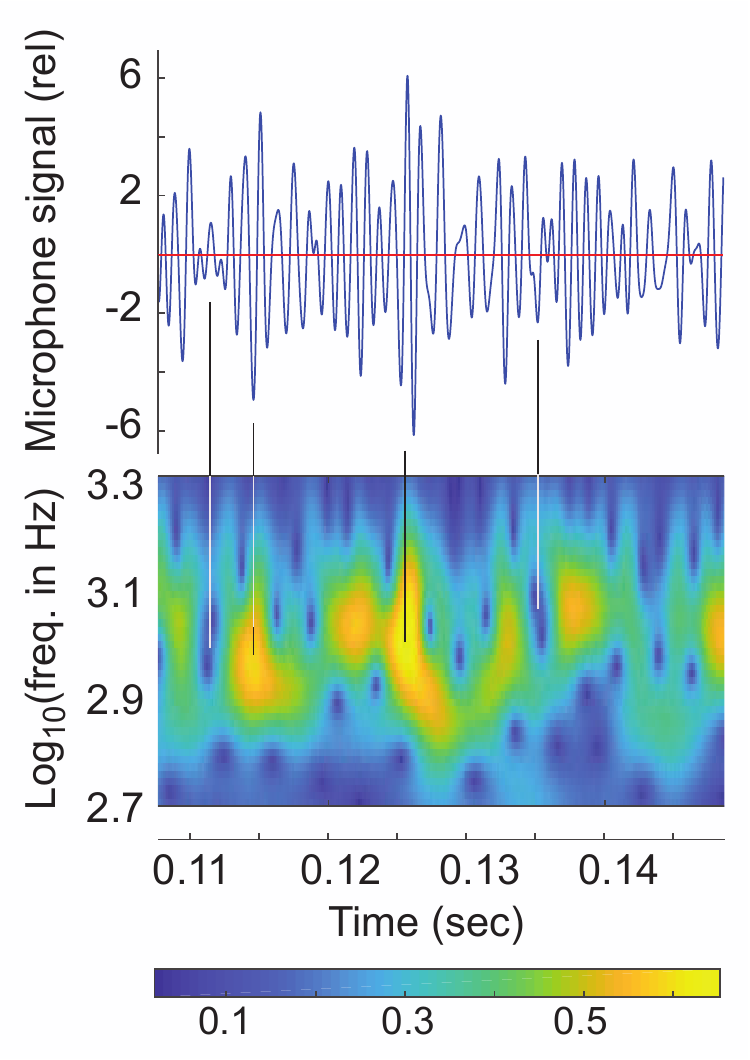} 
  \caption{Upper part: Amplitude vs time for a portion of a wave generated numerically (center frequency 1000 Hz, full bandwidth down to 1/e is 1000 Hz, sampling frequency 44.1\,kHz, only a 40 ms part of the signal is shown). Lower part: Time-frequency analysis (TFA diagram) for the same time interval for the frequency range 500 - 2000 Hz (log10 approx. 2.7 - 3.3). Light (yellow) patches correspond to time/frequency slots where the wave have a high amplitude within a particular frequency band. Dark (blue) areas indicate that the amplitude is moderate or low and no particular frequency is dominating within the corresponding frequency band (blue area in the diagram). The TFA diagram gives no information about the phases of the waves.}
  \label{fig:oneSignal}
\end{figure}

A little piece of such a wave is shown in the upper part of Fig.~\ref{fig:oneSignal}. It is an irregular train of more and less overlapping ``wavelets'' with a temporal duration and spectral width limited by the time-bandwidth product. The dominating frequency of each wavelet varies within the spectral width of the overall signal. This is characteristic for broadband waves. It is, however, difficult to get a kind of simple picture of the variations by using an amplitude vs time display alone (upper part of Fig.~\ref{fig:oneSignal}).

We found it very useful to visualize the statistical characteristics of broadband waves by a time-frequency analysis (TFA) (lower part of Fig.~\ref{fig:oneSignal}), In such a diagram high amplitude regions in time and frequency shows up as light patches (in our displays).

Time-frequency analysis is now available in a Matlab toolbox and in Python libraries. We chose the increasingly popular scalograms of the continuous wavelet transform (CWT) based on Morlet wavelets~\cite{LangForinash}, and used the version of this analysis described by Vistnes.~\cite{Vistnes}.

The relative frequency resolution is constant for all frequencies in the scalogram of the CWT type, based on Morlet wavelets. This is a nice feature since it makes it easier to optimize the choice of resolution in time vs resolution in frequency. (A parameter K gives approximately the number of oscillations in the wavelet, independent on the frequency. Low (high) K-values give high (low) time-resolution and low (high) relative frequency resolution.)

This detail is important since time and frequency are tightly connected by the time-bandwidth product ($\Delta f\Delta T\geq1$) where $\Delta f$ is the spectral width (in frequency space) and $\Delta T$ is the temporal duration of each wavelet. This is the classical analogy to Heisenberg's uncertainty principle. 

Within each light patch in the time-frequency diagram, it is possible to give an approximate frequency, phase and amplitude of the wave. However, all these parameters changes from one light patch to another. 

It should therefore, be pointed out, that it is impossible to define a global phase or a global frequency or wavelength for broadband waves. The summation of broadband waves leads to non-correlated results from time-slot to time-slot, which is quite different from the summation of harmonic waves or near harmonic waves that we usually meet in physics education. Each independent broadband wave is unique. 

The patchy pattern in the TFA-diagram works like a \emph{fingerprint of a broadband wave}. Every time we generate a new independent wave, the patchy pattern differ from the previous ones, even if the frequency distribution is identical. 

Two important tools can characterize a broadband wave: The autocorrelation function, and cross correlation between different broadband waves.

The autocorrelation function $ac$ is defined by
\begin{equation}
ac(\tau) = \frac{\sum_{t} {W(t) \cdot W(t + \tau)}}{\sum_{t} {W(t) \cdot W(t)}} 
\label{eq:acDef1}
\end{equation}

If we describe the wave for a total time much longer than the period of the dominating frequency components in $W$, the autocorrelation function is simply (See Eq. \eqref{corr} for a derivation):
\begin{equation}
ac(\tau) = \frac{\sum_{k=1}^{N} {|a_k|^2 \cos(\omega_k \tau)}}{\sum_{k=1}^{N} {|a_k|^2}}
\label{eq:acDef2}
\end{equation}

Fig. \ref{fig:CCprinc}a shows the autocorrelation function for our waves. The pattern is similar to a ``damped oscillations''  with a characteristic frequency (like the center frequency in our Gaussian frequency distribution) and a decay time of the order the inverse of the width of the frequency distribution.

The decay of the autocorrelation function tells us roughly how long time it takes from we have a complete knowledge of the wave until this knowledge is lost (except form the overall statistical description).

The autocorrelation function is exactly the result we get from a Michelson interferometer experiment. People that have performed a Michelson interferometry experiment on sunlight, will recognize that their result was rather close to the pattern seen in Fig. \ref{fig:CCprinc}a (but then both the positive and negative branch is seen).

On the other hand, there is no correlation between one broadband wave $W_1$ and an independent different broadband wave $W_2$, even if they have the same statistical distributions. The cross correlation between the waves is the important parameter:
\[
CC = \frac{\sum_{t} {W_1(t) \cdot W_2(t)}}{{\sqrt{\sum_{t} {W_1(t) \cdot W_1(t)}}} {\sqrt{\sum_{t} {W_2(t) \cdot W_2(t)}}}}
\]

If we follow these waves for a very long time compared to the relevant periods according to the frequency distribution ($t_{max} - t_{min} \gg 2 \pi/\omega_0$), it turns out that the cross correlation tends to zero. There is no correlation between the two waves! 

If we follow the waves for a shorter time, the calculated cross correlation is small and statistically distributed symmetrically around zero if we repeat the calculations repeatedly.

It should also be mentioned, that the result does not matter whether $W_1(t)$ and $W_2(t)$ are calculated at the same time $t$ or if one is time shifted compared to the other.

\section{Summation of broadband waves}\label{sum}

In HBT experiments, many non-correlated broadband waves add to each other at the positions of the detectors. Let us explore by numerical modeling the characteristic properties when we add two broadband waves to each other. In Fig.~\ref{fig:sumSignal} we show a tiny part of two non-correlated waves \#1 and \#2 generated with identical spectral width and center frequency (upper row). Note the difference in the fingerprint pattern! The lower row shows two different summations of wave \#1 and \#2 (details follow). The summation is interesting in several ways:

\begin{figure}[h] 
  \centering
  \includegraphics{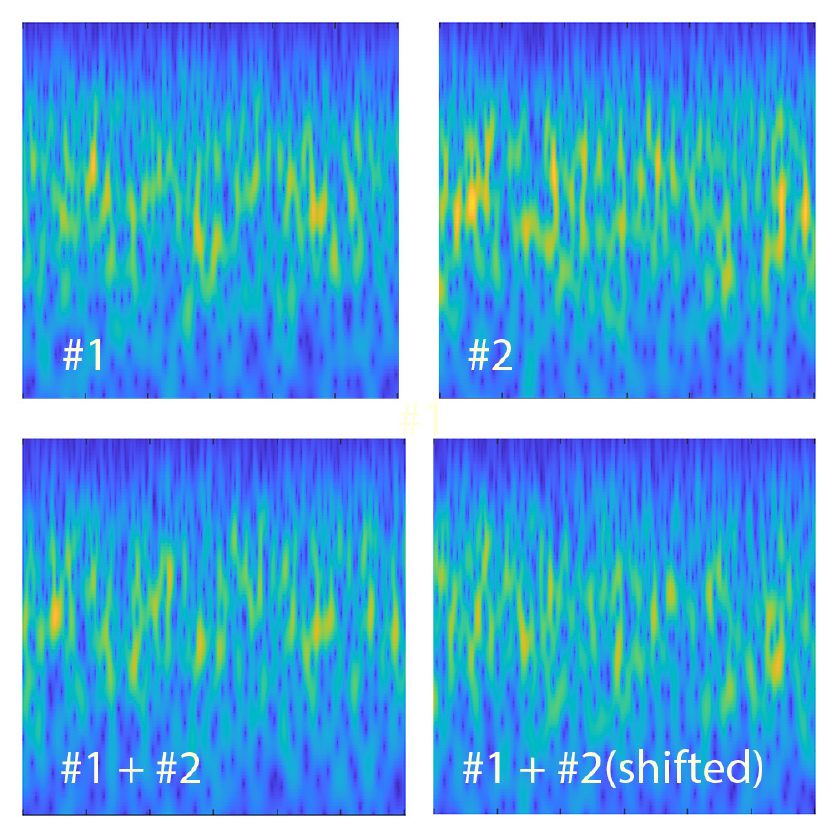}
  \caption{TFA diagrams for four different waves. The 0.3 s section and the frequency span along the y-axis are all the same. \#1 and \#2 are two different broadband waves with identical frequency distributions. The lower row shows summations of these two waves. In the lower right, wave \# 2 is time shifted 0.5 milliseconds (only 0.17 percent of the time span shown) before before the summation. The ``fingerprint'' pattern is clearly different for all four diagrams. See text for more details.}
  \label{fig:sumSignal}
\end{figure}

\begin{itemize}
\item{The mean amplitudes of the sum is approximately a factor $\sqrt{2}$ (not shown) as it should for broadband waves.}

\item{The TFA-diagram of the sum is qualitatively equal to each of the separate waves. The fingerprints are different, but it would still be impossible from the sum-TFA-diagram alone to judge if the wave was generated by one single source or from a sum of several sources.}

\item{More important in our connection is that the fingerprint pattern of a sum wave (lower row in the figure) is not a simple sum of the TFA-diagrams of the two original signals. The reason is that the two waves interfere constructively at some time-frequency sections, but destructively at some other time-frequency sections. The result depend on the detailed variation of frequency, amplitude and phase for each TFA section of the two waves in the sum.}

\item{The sum changes if the two waves are time-shifted relative to each other. The effect depends on the size of the time shift. If the time shift is much less than half the period $T_m$ of the dominating frequency, the sum wave is not changed much. However, the changes get more and more dramatic as the time shift gets close to $T_m/2$ and beyond. The effect is more complex than what we are used to for harmonic waves, and different for each of the light patterns in the TFA-diagram! More details follows in the next section.}

\end{itemize}

\section{Summation of waves in a typical HBT experiment}\label{princ-numeric}

\begin{figure}[h] 
  \centering
  \includegraphics{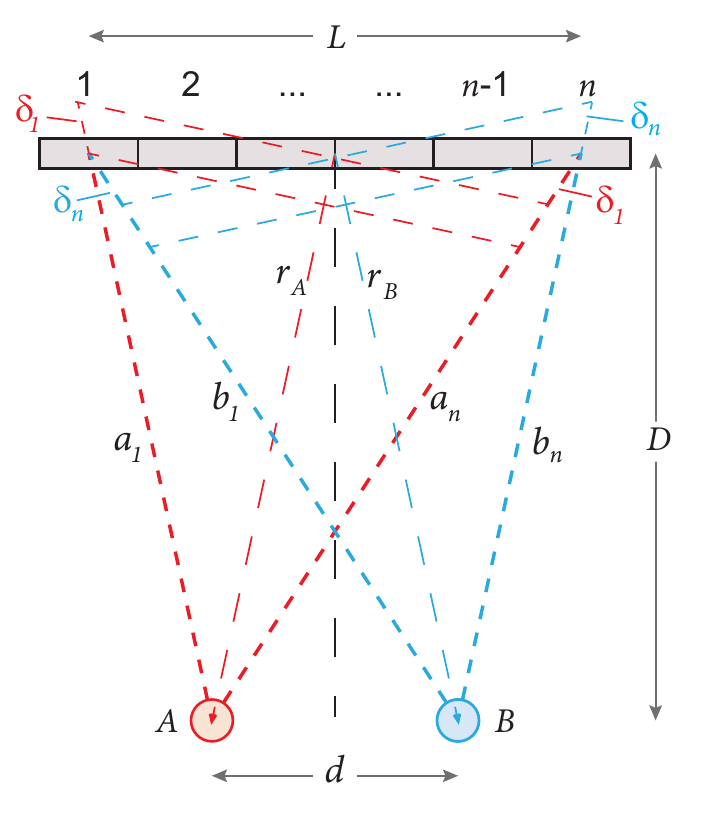}
  \caption{The principal geometry for an HBT experiment: A source consisting of a large number of independent broadband waves sources (here aligned along a line) and two detectors. The cross correlation between the detected signals is determined as a function of the distance between the detectors. For further details, see the text. }
  \label{fig:HBTgeometrics}
\end{figure}

We will now apply the basic knowledge about summation of broadband
waves to a HBT experiment. Typically, many (1,...,$n$) independent but
identical sources of broadband waves are then confined to a limited
area in space. The sources are positioned along a straight line with
length $L$ (see Fig.~\ref{fig:HBTgeometrics}). There are two identical
detectors, $A$ and $B$ placed a distance $D$ from the source line, in
a symmetrical manner as shown in the figure. The distance $d$ between
the detectors can be varied. Waves from all $n$ sources will hit both
detector $A$ and $B$, and the resulting wave is a sum of all these
part-waves at the positions of the detectors.

We will use the symbol $W_A (t)$ for the wave at detector $A$ at the time $t$, and $W_1(t)$ the wave emitted from source 1 at time $t$. We assume that $D \gg d$ and $D \gg L$, thus we neglect a common reduction of amplitude due to the distance from source to detectors as well as differences in amplitudes whether one part-wave hit the detector $A$ or detector $B$.

The path-length for the wave from source 1 to detector $A$ is $a_1$ (see Fig.~\ref{fig:HBTgeometrics}). The time the wave spend from source element 1 to detector A is $t_{A,1} = a_1/c$ where $c$ is the speed of the wave. Similar nomenclature for the other path-lengths and detector combinations. 

We get for the summation:
\begin{align*}
W_A (t) &= W_1(t-a_1/c) + \cdots + W_n(t-a_n/c)   \\
        &= W_1(t-r_A/c+\delta_1/c) + \cdots + W_n(t-r_B/c-\delta_n/c)
\end{align*}
\begin{equation}
  W_A (t) = W_1(t'+\delta_1/c) + \cdots + W_n(t'-\delta_n/c)
  \label{eq:A}
\end{equation}
and
\begin{equation}
  W_B (t) = W_1(t'-\delta_1/c) + \cdots + W_n(t'+\delta_n/c).
  \label{eq:B}
\end{equation}
where $\delta$, $r_A$ and $r_B$ are defined by Fig.~\ref{fig:HBTgeometrics}, and we have used the symmetry $r_A = r_B$.

In a HBT experiment, the cross-correlation between the sum-waves at the detector $A$ and the sum-waves at detector $B$ is determined as a function of the distance between the detectors. The cross-correlation is defined by:
\begin{equation}
CC = \frac{\sum_{t} {A(t) \cdot B(t)}}{{\sqrt{\sum_{t} {A(t) \cdot A(t)}}} {\sqrt{\sum_{t} {B(t) \cdot B(t)}}}}
\label{eq:defcc}
\end{equation}
Since $A$ and $B$ are described by Eq.~\ref{eq:A} and \ref{eq:B}, the cross correlation will consist of $n^2$ terms, $n^2-2n$ of them will represent cross-correlations between \emph{different} part-waves. Since these part-waves are independent, all these $n^2-2n$ terms will be zero. The $2n$ terms that will survive in the numerator are:
\begin{align*}
&W_1(t'-\delta_1/c) \cdot W_1(t'+\delta_1/c) + W_2(t'-\delta_2/c) \cdot W_2(t'+\delta_2/c) + \cdots \\
&+ 2 W_{n-1}(t'+\delta_{n-1}/c) \cdot W_{n-1}(t'-\delta_{n-1}/c) + 2 W_n(t'+\delta_n/c) \cdot W_n(t'-\delta_n/c)
\end{align*}
We notice that for the sum-wave that hits detector A, the wave from source 1 is time-shifted before the wave from source $n$, while it is the other way around for the detector B. Thus, from Fig.~\ref{fig:sumSignal} it follows that the TFA-diagrams from detector A and B will not differ significantly only if $\delta_1/c$ is much less than half the period of the dominating wavelength, $T_m/2$.

If the statistical properties of the waves does not change over time, the cross-correlation will, according to Eqs.~\ref{eq:acDef1} and \ref{eq:acDef2}, be proportional to \emph{a sum of autocorrelation values}:
\[
\frac{1}{N'}\left(ac(2\delta_1/c) + ac(2\delta_2/c) + \cdots + ac(2\delta_{n-1}/c) + ac(2\delta_n/c)\right)
\]
However, $\delta_1/c = \delta_n/c$ and the autocorrelation function is symmetrical arround zero. Thus, by choosing $n$ as an even number, the cross-correlation will be proprtional to
\begin{equation}
CC \propto (ac(2\delta_1/c) + ac(2\delta_2/c) + \cdots + ac(2\delta_{n/2}/c)
\label{eq:ccAsSumac}
\end{equation}
where $\delta_{n/2}/c \approx 0$ for large number of our description of the sources of part-waves.

The denominator in our cross-correlation calculation does not depend on the size of the $\delta$-s. In our experiments we will choose a normalization constant so that $CC = 1.0$ when the detectors are very close to each other.

If the $\delta_1$, $\delta_2$, $\cdots$, $\delta_{n/2}$ follow a linear curve from a maximum to zero, the cross-correlation between the sum-waves at detector $A$ and the sum-waves at detector $B$ is simply the mean value of the autocorrelation function of the part-waves from zero to a maximum value given by $2\delta_1$. We denote this limit 
\begin{equation}
\Delta = 2\delta_1
\label{eq:defDelta}
\end{equation}
$\Delta$ is then the maximum difference in distance between a detector (e.g. $A$) to the various positions in the HBT array of sources of waves.

\begin{figure}[h] 
  \centering
  \includegraphics{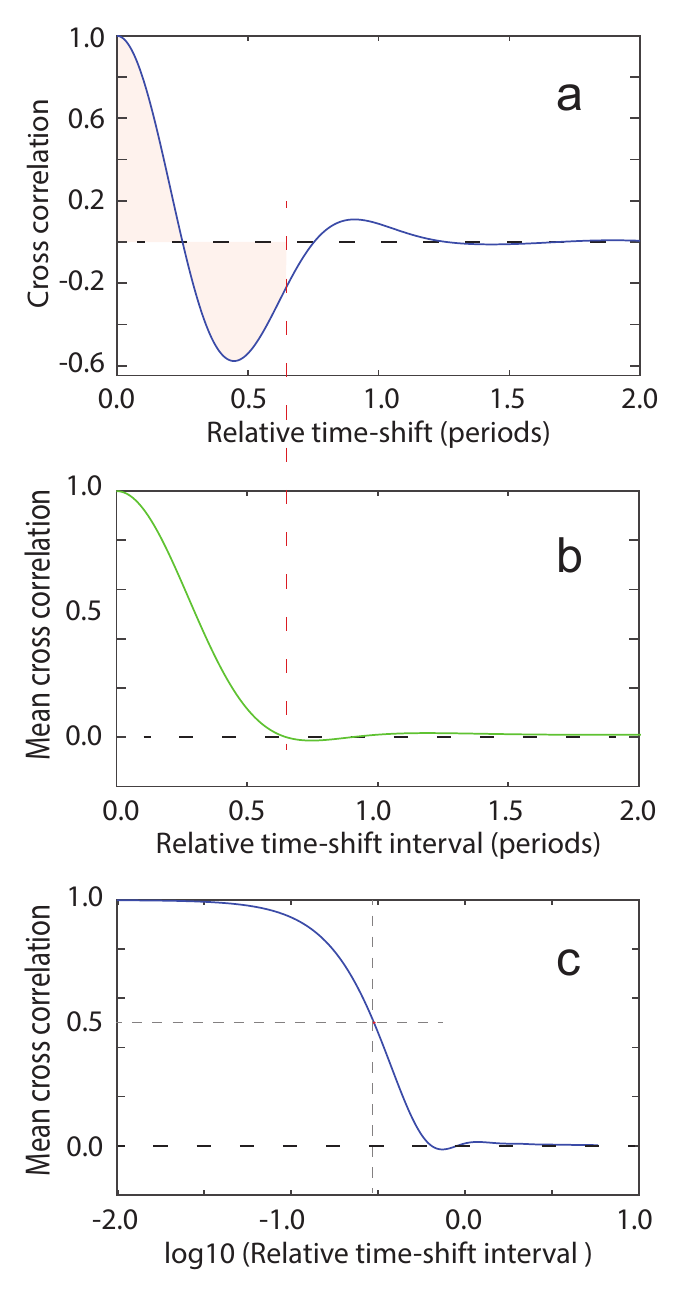}
  \caption{a) The autocorrelation function for our modelled broadband waves. The x-axis is given in proportions of the period of the dominating frequency. b) The cross-correlation of the waves at detectors $A$ and $B$ is according to Eq.~\ref{eq:ccAsSumac} the ``mean value'' of the autocorrelation function for an interval starting from zero and an increasing width. c) The very same curve as in b), but drawn with time-shifts given in a logaritmic (log10) axis. The mean autocorrelation value drops to 0.5 when $\Delta$, the maximum difference in distances between a detector and the sources, is 0.3 times the dominating wavelength (log10(0.3) = -0.52). This figure will be compared with similar figures for the experimental and theoretical data later in this paper.}
  \label{fig:CCprinc}
\end{figure}

Knowledge of how the cross correlation changes with the distance between the detectors makes it possible to calculate the ratio $L/D$, or the angle of the entire source of waves, if we know the wavelength of the dominating frequency of the waves. This was the basic feature in the original HBT experiment. We only want to demonstrate that our numeric and analytical descriptions actually can experimentally be verified in a complete other setting than originally.

\section{Experimental}

We will now present an experimental realization of the model used in our numerical presentation. We will also show the deep connection between analyses at the amplitude level and intensity level.

We chose to use sound waves where the frequency is far less than the sampling frequency of readily available inexpensive digitizers. This allows us to start at the amplitude level and add analysis at the intensity level as well. 

\begin{figure}[h] 
  \centering
  \includegraphics{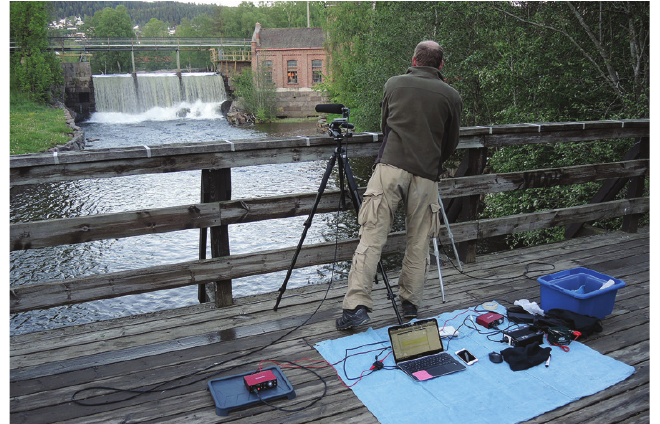}
  \caption{Experimental setup with two microphones directed towards the center of the waterfall. }
  \label{fig:expPhoto}
\end{figure}

\begin{figure}[h] 
  \centering
  \includegraphics[width=4cm]{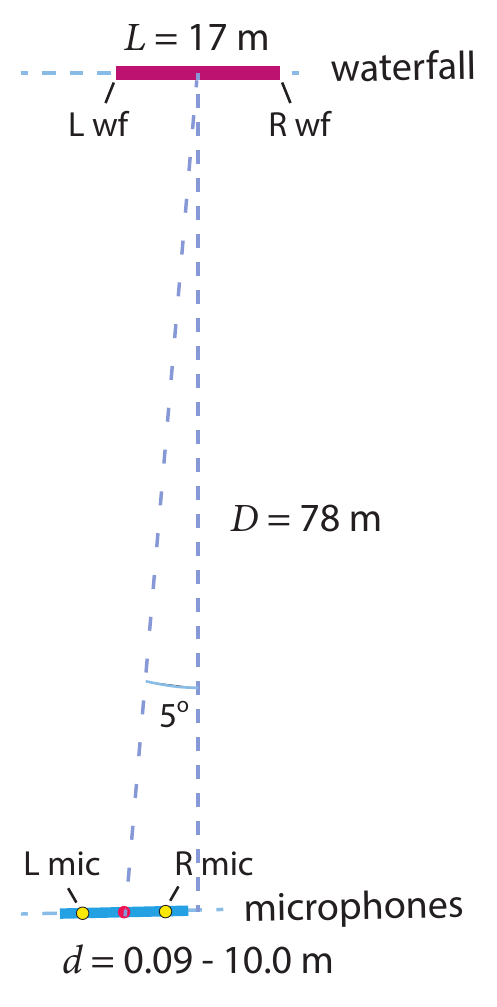}
  \caption{Geometrical relationship between the waterfall and the line of positions for the microphones. A red dot indicates the symmetry point for the microphones. See the text for more details.}
  \label{fig:stilla}
\end{figure}

A waterfall at a dam was a suitable source of waves. 
%Akerselva (river) located approximately 300 m west of the Norwegian Museum of Science and Technology in Oslo, Norway.
The waterfall has a relatively even water-flow across all of its width (approx. 17 m) (Figs.~\ref{fig:expPhoto} and \ref{fig:stilla}). We measured the sound from the waterfall by two microphones mounted on camera tripods. The distance between the microphones varied between 0.09 and 10 m, symmetrically around a midpoint. We placed the front of the microphones along a rail at a walking bridge almost parallel to the waterfall, at a distance approximately 78 m from the waterfall. The microphones pointed directly to the center of the waterfall during measurements. The sound level from the waterfall was approximately 60 dB(A). There were no obvious other sources of sound in the neighborhood since we carried out the measurements very early in the morning before the traffic started on the nearby road. The landscape on both sides of the waterfall had no obvious sound reflecting objects, thus the recorded sound was mainly direct sound from the waterfall.

Sennheiser MKH 8070 microphones were used. They have a strong directivity in the sensitivity, but even so an almost perfect uniform sensitivity throughout the angular extension of the waterfall (13 degrees). It was less than 1-2 dB difference in sensitivity for sound depending on which part of the waterfall the sound came from.

Two Focusrite Scarlett Solo 2nd generation audio interfaces powered and amplified the microphone signals. A National Instruments USB 6211 16 channel 16 bit max 250 kS/s multifunction I/O unit digitized the resulting signals, and a Dell XPS laptop run the digitizer at a sampling rate of 44150 S/s. The time difference between the exact sampling for the two microphone signals was so small that it only influences results for higher sound frequencies than we are using in this work.  The data sampling lasted 5 s, and the measurements were repeated three times for each chosen microphone distance.

All basic power to the electronic measurement equipment came from a Mascot inverter Type 2285 connected to a 12 V battery. The inverter provided 230 V true sine wave to minimize the noise from this source. We found no trace of particular 50 Hz harmonics in the recorded signal. 

We wrote the software for analyses of the results in Matlab. Essential functions were sampling and storing to disk, fast Fourier transform, frequency filtering based on Fourier transform, calculation of autocorrelation function, cross correlation, and the same time-resolved frequency analysis as presented above (based on Morlet wavelets as described elsewhere~\cite{Vistnes}).

As discussed in the numerical modeling section, our analysis was based on the assumption that it is a linear distribution of differences in distances between a detector (e.g. $A$) and the various positions along the waterfall (the $\delta_k$-values).

A closer look reveals that the distances $a_1$, $a_2$, $\cdots $,$a_{n-1}$, $a_n$ follow a parabolic curve, but the \emph{differences} in distances $a_1-a_n$, $a_2-a_{n-1}$, $\cdots $ that is important for us, turn out to follow a linear curve. Thus, the analysis based on mean values of the autocorrelation function illustrated in Fig.~\ref{fig:CCprinc} is valid for our experimental situation.

\section{Results, analysis of the correlations at the amplitude level}

The sound waves that hit the microphones are the sum of broadband sound waves coming from all over the waterfall. The signals from the two microphones are proportional to the amplitudes of the sum-waves at the positions of the microphones. It is the geometry that determines the summation and leads to different sum-waves at the microphones as the distance between them is increased. We use a geometry very close to a standard HBT setup (compare Fig.~\ref{fig:HBTgeometrics} and Fig.~\ref{fig:stilla}).

\begin{figure}[h] 
  \centering
  \includegraphics{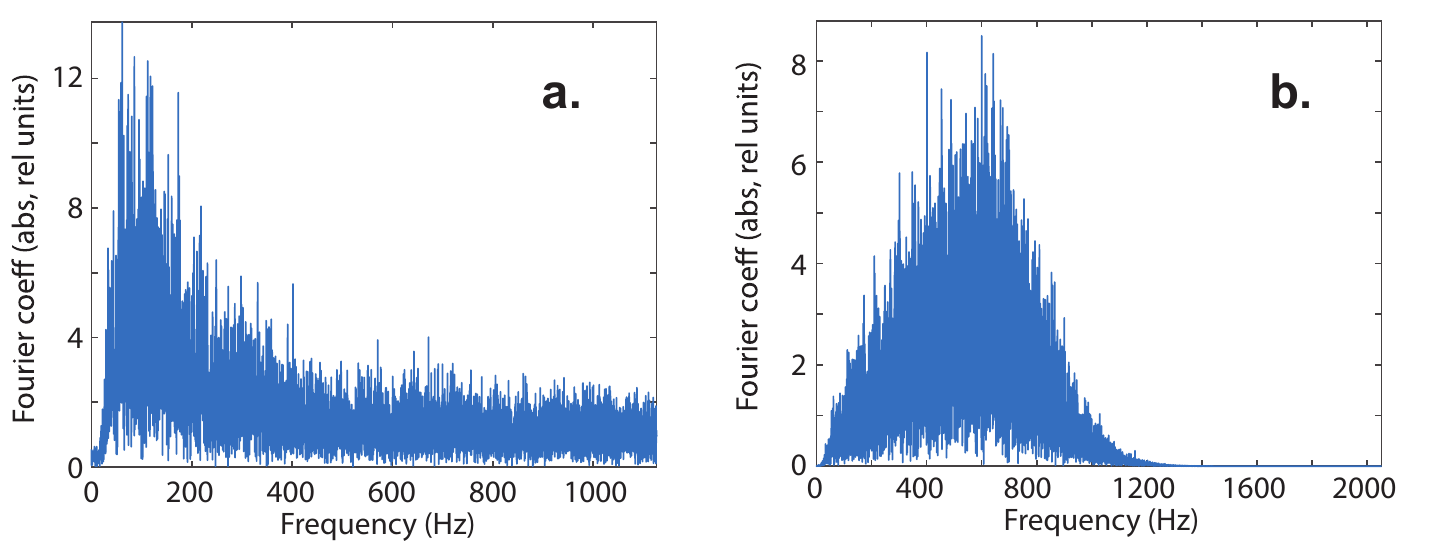}
  \caption{a. Part of the full frequency spectrum of the sound from the waterfall. b. Frequency spectrum of the original signal filtered with a Gaussian filter with center frequency 600 Hz and full width 600 Hz. The asymmetry in the final frequency spectrum is due to the peak between 100 and 200 Hz in the original signal.}
  \label{fig:sound}
\end{figure}

The frequency spectrum of the sound recorded at the amplitude level showed a broad peak around 95 Hz. The intensity stretched from approximately 25 to 8000 Hz. Fig. \ref{fig:sound}a shows only a part of the frequency spectrum.

The very broad frequency spectrum makes it impossible to point out one dominating frequency. We therefore performed a frequency filtration with a Gaussian distribution function, based on fast Fourier transform (FFT) and inverse FFT after the filtering. This allowed us to use different frequency bands in our analysis, and our choices were:

\begin{center}
\begin{tabular}{ c c }
 Center frequency & Full bandwidth \\ 
 (all frequencies in Hz) & (down to 1/e of max) \\
 100 & 100  \\  
 300 & 300  \\ 
  600 & 600  \\ 
   1000 & 1000  \\ 
\end{tabular}
\end{center}

Fig. \ref{fig:sound}b shows an example of the frequency distribution for a filtered signal. 

\begin{figure}[h]
  \centering
  \includegraphics{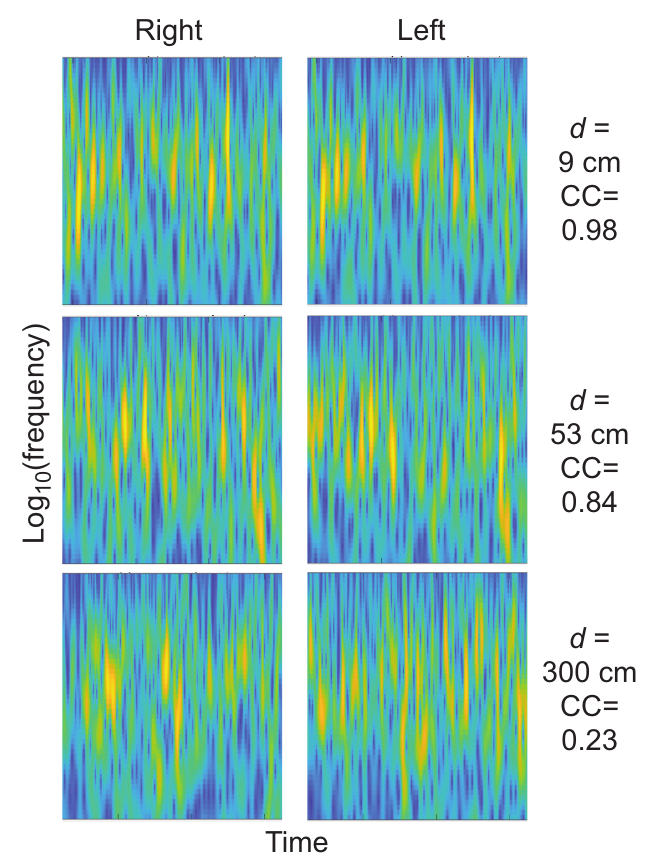}
  \caption{Time-frequency analysis for 600 Hz filtered microphone signals for three chosen distances between the microphones. The pattern is close to equal when the microphones are only 9\,cm apart (upper part), slightly different for 53\,cm distance between the microphones (middle), and vastly different when the microphones are 300\,cm apart (bottom). This correlates with the cross correlations (CC) calculated for each set of $\approx$5 s recordings. In all the six TFA-diagrams the time interval shown is 40 ms (corresponding time slot for each set) and the frequency range is 350-950 Hz. The microphone signals were filtered by the 600 Hz filter mentioned in the text. In the TFA-diagrams the K-value (discussed above) is 6 (high time resolution, moderate frequency resolution).}
  \label{fig:treAvst}
\end{figure}

As pointed out in the numerical descriptions above, a time frequency analysis can ease the understanding of the physical process behind the HBT effect, and Fig.~\ref{fig:treAvst} gives an example. Note the logarithmic scale on the frequency axis. The frequency span is 300 - 1100 Hz. These diagrams are similar to the numerical modeling in Fig.~\ref{fig:sumSignal}.

\begin{figure}[h]
  \centering
  \includegraphics{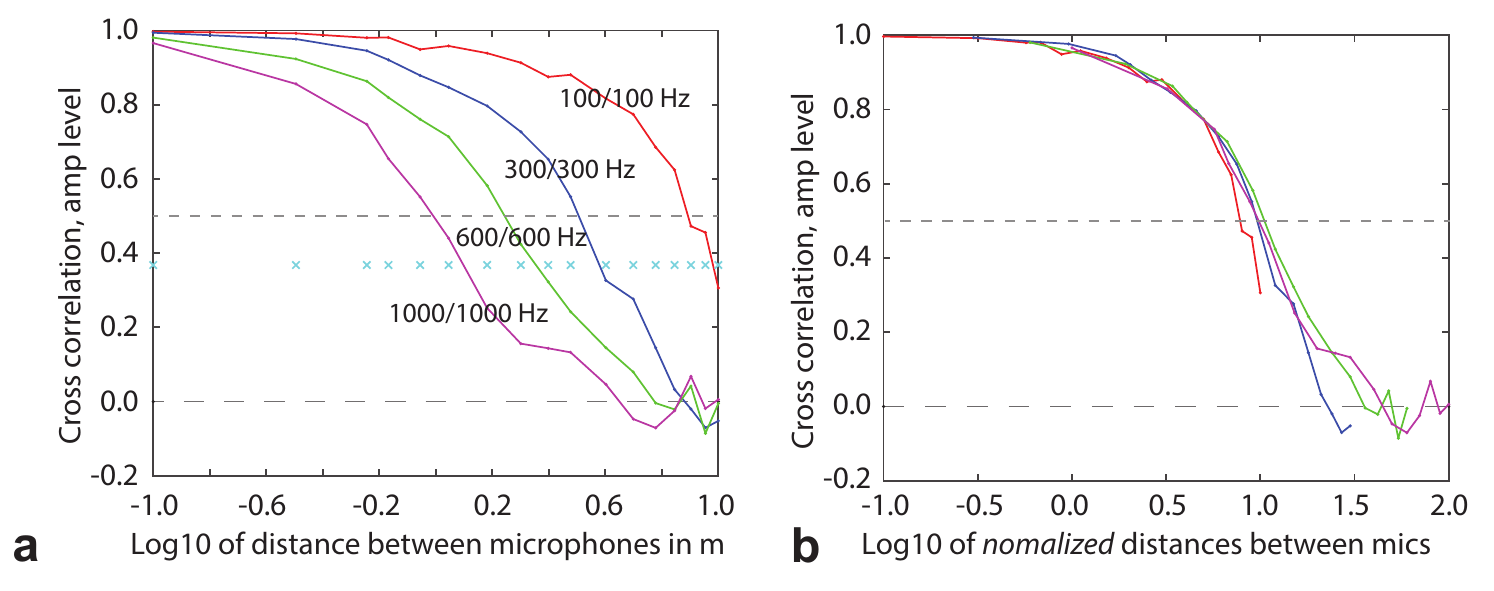}
  \caption{a. Cross correlation between the two microphone signals as a function of distance between the microphones. The numbers in black are the four different frequency bands used in the analysis. Blue +-signs indicate the actual distances between the microphones. Each point in the graph is the mean value for three recordings. b. The very same data, but the x-axix is 'normalized distances' between the microphones (see text).}
  \label{fig:ccAmpl}
\end{figure}

As discussed above, the correlations between the patterns in the TFA-diagrams are more accurately determined by the cross correlation calculations (CC) between the two microphone signals. Figure \ref{fig:ccAmpl}a shows how the cross correlation changes with the distance between the two microphones (a logarithmic scaling is used) for the four different frequency bandwidths.

Since we use cross correlations between microphone signals, our experiment can be characterized as a kind of multiplicative interferometry. However, since the waves at the positions of the microphones actually is a sum of waves at the amplitude level, it is also a kind of additive interferometry inherent in the experiment.

We notice that the cross correlation seems to correspond very well with the results from our numerical modeling in Fig.~\ref{fig:CCprinc}. We will return to some more details below.

A difficulty should here be mentioned. The five degree asymmetry in the experimental geometry in Fig.~\ref{fig:stilla} compared to the ideal situation in Fig.~\ref{fig:HBTgeometrics} leads to a systematic difference in $r_A$ and $r_B$ i Fig.~\ref{fig:HBTgeometrics}. We had to compensate for this asymmetry, by a systematic and progressive larger time shift of signal A relative to signal B before the cross correlations were calculated as microphone distance increased. Details are available from the authors.

The curves with cross correlations in Fig.~\ref{fig:ccAmpl}a start close to 1.0 when the microphones are close to each other, but after a characteristic plateau it drops relatively abruptly down to about zero. As a parameter to describe the differences between the different curves, we use the microphone distance that correspond to a cross correlation of 0.5 (similarly as described in Fig.~\ref{fig:CCprinc}). These distances are approximately 8.1, 3.2, 1.7 and 1.0 m for the filter frequencies of 100, 300, 600 and 1000 Hz.

We note that shapes of the four curves are similar (for logarithmic scaling of the microphone distances). We also note that as the filter frequency increases (wavelengths decreases) there is a shift towards shorter microphone distances.

In Fig.~\ref{fig:ccAmpl}b we have used ``normalized'' distance between the microphones, where the actual microphone distance is multiplied by 3, 6 and 10 for the data we obtain when the analysis frequency is 300, 600 and 1000 Hz, respectively. Thus, the different curves all correspond to an ``effective'' wavelength equal to the wavelength for 100 Hz filtering.

The four curves seem to more or less overlap, indicating that the HBT effect relate closely to the wavelength of the dominating wavelength of the sum signal. A detailed explanation for the collapse of the scaled curves will be given in Section~\ref{theory}.

\section{The crucial connection between amplitude level and intensity level}

So far, our analyses have been at the amplitude level for the waves. The original HBT effect was based on analyses at the intensity level. Let us look into the connection between these levels of detection and analysis. 

The intensity of a wave is proportional to the amplitude squared. If we just choose two components in the summation of waves, we have (at an arbitrary position in space):

  	Amplitude level:    
  	\[
  	f(t) = A \cos(\omega_1 t + \phi_1) + A \cos(\omega_2 t + \phi_2)
  	\]
 
	Intensity level: $I(t) = f^2(t)$: 
	\begin{align} 
	I(t) =   
	& \, A^2 \left( \cos^2(\omega_1 t + \phi_1) + 2\cos(\omega_1 t + \phi_1) \cos(\omega_2 t +
			 \phi_2) + \cos^2(\omega_2 t + \phi_2) \right) \nonumber \\
		     = & \, A^2 \left( \cos((\omega_1 + \omega_2) t + \phi_1+\phi_2) + 
			\cos((\omega_1 - \omega_2) t + \phi_1-\phi_2)\right) \nonumber \\
			+ & \, A^2 \left( \frac{1}{2} \cos(2 \omega_1 t + 2 \phi_1) + \frac{1}{2} \cos(2 \omega_2 t + 
			2 \phi_2) + 1 \right)
	\end{align}

The result is that the intensity signal will have frequency contributions both at the sum frequency and at the difference frequency.

The difference frequencies for a broadband signal can go all away down to zero frequency - it may in general even show up at frequencies several decades lower than the original signal.

\begin{figure}[h] 
  \centering
  \includegraphics{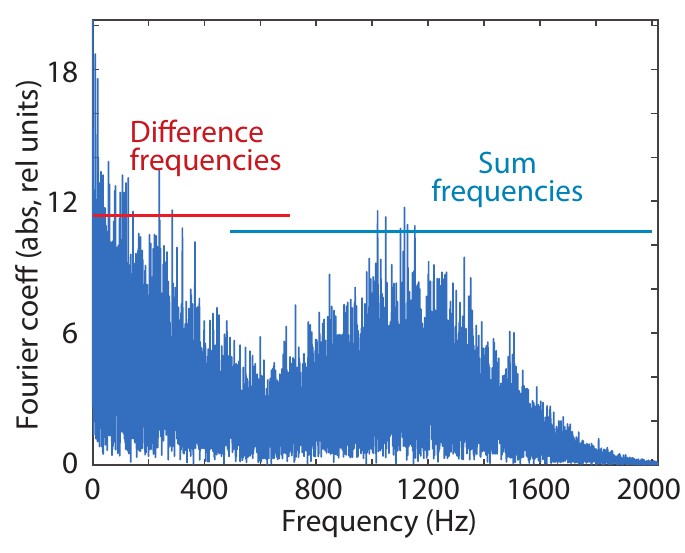}
  \caption{Frequency spectrum of the square of our microphone signal filtered with center frequency 600 Hz (same signal as given in Fig.~\ref{fig:sound}b). We removed the very high intensity point at zero frequency in this plot (more details in the text).}
  \label{fig:freqSquared}
\end{figure}

In our experiments, we recorded very broadband signals and filtered them to obtain a more limited frequency distribution. From these signals, the relative intensities are calculated by squaring the amplitude at every point in time. A Fourier analysis of a typical result is given in Fig.~\ref{fig:freqSquared}

Since the original signal in this example was filtered with a Gaussian filter centered at 600 Hz and the full bandwidth down to 1/e was 600 Hz (see Fig.~\ref{fig:sound}b), the frequency distribution of the intensity is very wide. There is an overlap between the difference and sum frequencies (see Fig.~\ref{fig:freqSquared}). If the original filter had been narrower, the difference and sum frequencies would have been more distinct.

In the HBT case, they used a passband filter when they analyzed the signal from the intensity detector. The passband frequency was several orders of magnitude lower than the frequency of the light. Even so, the passband frequency was so high that they could measure the cross correlations within an acceptable recording time.

In our case, we will calculate cross correlations for the complete intensity signal given by the mathematical expression above, including both difference- and sum-frequencies. We will in addition use a passband filter for picking out parts of the difference frequency signal to study how this influences the result.

\subsection{Analysis of the complete intensity signal}

\begin{figure}[h] 
  \centering
  \includegraphics{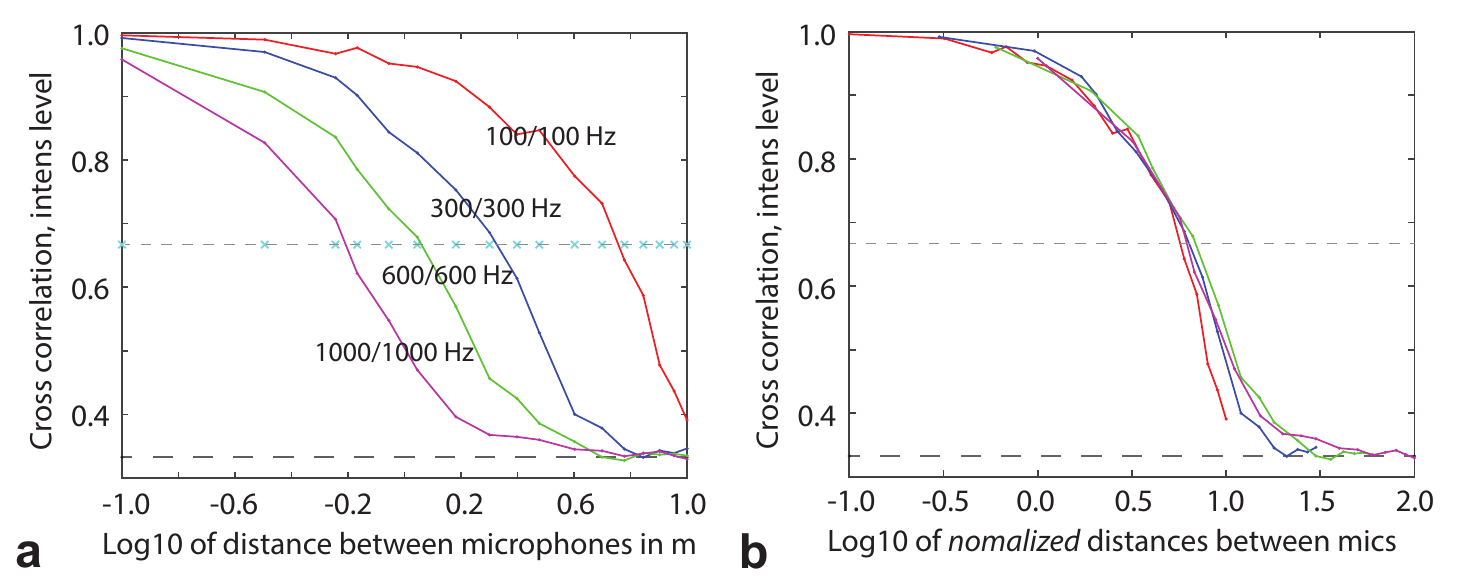}
  \caption{a. Cross correlation between the complete intensity signals at the two microphones as a function of distance between the microphones. Numbers in black indicate the different frequency bands. Each point in the graph is the mean value for three recordings. b. The very same data, but the x-axix is 'normalized distances' between the microphones. }
\label{fig:ccInt}
\end{figure}

In this case, we first filtered the microphone signal at the amplitude level using the same Gaussian filters as above. Squaring the amplitudes point for point along the complete recording gave us the intensity signals. Finally, we calculated the cross correlations between the right and left microphone intensity signals for all the 17 different microphone distances as before, and the results are given in Fig.~\ref{fig:ccInt}a.

The result is very similar to the result from analyses at the amplitude level in Fig.~\ref{fig:ccAmpl}a, but the curves are slightly shifted towards smaller microphone distances. We also notice that the curves do not go to zero for large microphone distances, but to a level that seems to be roughly 0.33, see Eq.~\ref{intensityNorm} below for an explanation for this. Thus, instead of choosing the 0.5 cross correlation as a parameter to characterize the drop from full cross correlation to no cross correlation as we did at the amplitude level, we here chose the 0.67 cross correlation points as a characteristic parameter.

The microphone distances that correspond to this 0.67 cross correlation points are then 5.8, 2.1, 1.2 and 0.6 m for the signals with dominating frequency 100, 300, 600 and 1000 Hz, respectively.

Again, the shifts of the curves in Fig.~\ref{fig:ccInt}a indicate a wavelength dependency, and this is verified by Fig.~\ref{fig:ccInt}b where we use the ``normalized'' distances between the microphones (see above). In this figure, the four curves overlap each other quite well.

\subsection{Analysis of a low-pass filtered intensity signal}

In the original HBT work, analysis were performed at a frequency band very much lower than the frequency of the light itself. In our case, it is difficult to obtain so distinct frequency bands for the original signal compared to the analyzing signal, since the frequency of the sound from the waterfall is low. However, we played around with various analyzes where the full intensity signal was filtered before the calculations of the cross correlations to mimic the process used in the original HBT work. The cross correlation parameter for these filtered intensity signals came out equal to the full intensity signal (within the uncertainty level, data not shown). \emph{This is important, since it tells us that we would have gotten the same results even if we had to rely only on intensity fluctuations at a frequency band far lower than the frequency of the waves itself (as was the case in the original HBT work)}.

\section{Theoretical treatment of our system}\label{theory}

In this section we provide a theoretical analysis of the experiment, showing how one can explain the scaling relation between different filters as demonstrated in Figs.~\ref{fig:ccAmpl}b and \ref{fig:ccInt}b as well as the fact that the intensity correlation in Fig.~\ref{fig:ccInt} decays to a level different from zero. The model that we use is a direct generalization of the large number of sources model presented in Fig.~\ref{fig:HBTgeometrics}.

\begin{figure}[h] 
\begin{center}
  \includegraphics{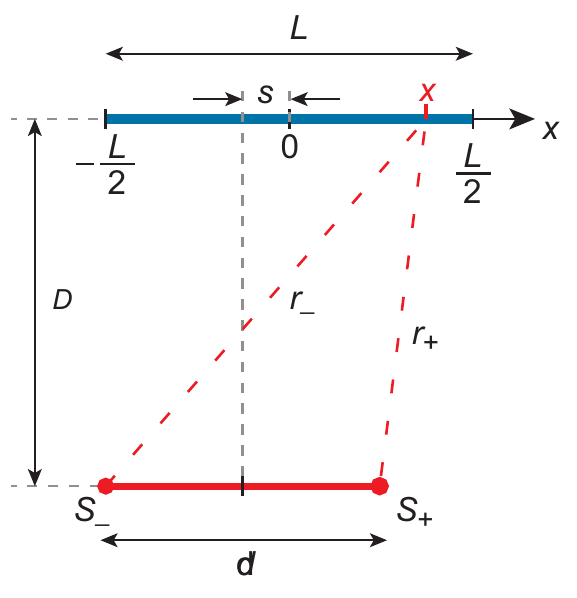}
  \caption{Symbols used in the theoretical analysis.}
  \label{fig:theory}
\end{center}
\end{figure}

We consider (see Fig.~\ref{fig:theory}) a line of sources with length $L$. The detectors are separated by a distance $d$ and the distance from the source line to the midpoint between the detectors is $D$. The midpoint of the line of sources is displaced a distance $s$ from the center line between the detectors. (Thus, the actual asymmetry in our experiment is here taken into consideration).
  
We assume that there are independent sources of broadband waves at each position $x$ and that these waves hit the detectors with approximately equal amplitude but different time shifts. A wave emitted at $x$ at time $t$ leads to a signal $f(x,t')$ when the wave hit a detector at time $t'$. The signal received at the two detectors is then (assume linear detectors)
\[
S_\pm(t) = \int_{-L/2}^{L/2}dx f(x,t-r_\pm/c)
\]
where $c$ is the speed of sound and $r_\pm$ are the distances from the source at $x$ to the two detectors. We then have

\begin{equation}
r_\pm = \sqrt{D^2+(x+s\mp\frac{d}{2})^2}\approx D+\frac{1}{2D}(x+s\mp\frac{d}{2})^2
\label{eq:rpm}
\end{equation}
when $D$ is large compared to $x$, $s$ and $d$.
The cross correlation of the amplitudes in the two detectors is
\[
\langle S_+S_-\rangle = \int dx_1 dx_2\langle f(x_1,t-r_+(x_1)/c)f(x_2,t-r_-(x_2)/c)\rangle
\]
We assume that the sources at different $x$ are uncorrelated and stationary so that
\[
\langle f(x_1,t_1)f(x_2,t_2)\rangle = \delta(x_1-x_2)C(t_1-t_2)
\]
where $C(t)$ is the autocorrelation function for the signal from a single source (we assume that all sources are identical). This is an anology to our numerical modeling above. Using Eq.~\ref{eq:rpm} we get 
\[
 \langle S_+S_-\rangle = \int_{-L/2}^{L/2} dx C(r_-(x)/c-r_+(x)/c) =  \int_{-L/2}^{L/2} dx C\left(\frac{d}{Dc}(s+x)\right)
 \]
 
Let us now assume that the sources are broadband with a large number of frequencies contributing. Then the source will produce a signal
 \[
 f(x,t) = \int d\omega A(\omega)\cos(\omega t+\phi_\omega)
 \]
The amplitudes $A(\omega)$ either describe the actual spectral distribution of the source, or the characteristics of the filter, as we will use below. Note that $A(\omega)$ in principle also depends on $x$. That is, each source can have a different spectrum. We will assume that all sources are statistically equivalent, and that the resulting autocorrelation function is independent on $x$. The phases $\phi_\omega$ are random numbers between 0 and $2\pi$ and independent for each $\omega$.  This gives the correlation
 \[
 C(\tau) = \frac{1}{2T}\int d\omega d\omega'A(\omega)A(\omega')
   \int_{-T}^Tdt\cos(\omega t+\phi_\omega)\cos(\omega'(t+\tau)+\phi_{\omega'})
 \]

  The average of the cosines is zero, unless $\omega'=\omega$, and we get 
 \begin{equation}\label{corr}
  C(\tau) = 
 \frac{1}{2}\int d\omega A^2(\omega)\cos\omega\tau.
 \end{equation}
 
 With a Gaussian filter
 \[
 A(\omega) = e^{-\frac{(\omega-\omega_0)^2}{2\sigma^2}}.
 \]

 \[
 C(\tau) = \frac{1}{2}\int d\omega  e^{-\frac{(\omega-\omega_0)^2}{\sigma^2}}\cos\omega\tau  = \frac{\sqrt{\pi}\sigma}{2}e^{-\frac{\sigma^2\tau^2}{4}}\cos\omega_0\tau.
 \]
 We then get

   \begin{equation}\label{CCAmp}
     \langle S_+S_-\rangle = \frac{Dc}{ Ld\omega_0} 
   \int_{-u_0+u_s}^{u_0+u_s}du e^{-\alpha u^2}\cos u.
   \end{equation}
   where we have  normalized  by requiring that $\langle S_+S_-\rangle = 1$ when $d=0$
and introduced the new variables

 \begin{equation}\label{x}
   u =\frac{d\omega_0}{Dc}(x+s) ,\quad u_0 =\frac{Ld\omega_0}{2Dc},\quad
   u_s =\frac{sd\omega_0}{Dc}\quad\mbox{and}\quad \alpha = \frac{\sigma^2}{4\omega_0^2}.
 \end{equation}

We filtered the signals with Gaussian filters before cross correlation was calculated. The center frequencies are $\omega_0/2\pi=$ 100 Hz, 300 Hz, 600 Hz and 1000 Hz, with $\sigma=\omega_0/2\sqrt2$ in all cases, which is the same as we used in the analysis of the observed sound spectra. From \eqref{CCAmp} and \eqref{x} we see that the cross correlation $\langle S_+S_-\rangle$ only depends on the frequency $\omega_0$ in the product $d\omega_0$, and that as long as we keep the ratio $\sigma/\omega_0$ fixed, $\alpha$ will be the same for all $\omega_0$. This means that we expect the four curves above to fall on the same curve if we use the scaled axis $d\omega_0$, as was shown in Fig.~\ref{fig:ccAmpl}b.

\section{Analysis at the intensity level} % Er dette en egnet overskrift???

The intensity is the square of the amplitude, $I_\pm = S_\pm^2$. The cross correlation of the intensities is
\[
\langle I_+I_-\rangle =  \int dx_1 dx_1' dx_2 dx_2'\langle f(x_1,t-r_+(x_1)/c) f(x_1',t-r_+(x_1')/c)f(x_2,t-r_-(x_2)/c)f(x_2',t-r_-(x_2')/c)\rangle
\]
Since the sources are independent at each position, the coordinates must be pairvise equal.
\[
  \begin{array}{l}x_1=x_1'\\x_2=x_2'
  \end{array}\qquad\mbox{or}\qquad
 \begin{array}{l}x_1=x_2\\x_1'=x_2'
  \end{array}\qquad\mbox{or}\qquad
 \begin{array}{l}x_1=x_2'\\x_1'=x_2
  \end{array}
 \]
 The last two give equal contributions because of symmetry and we get

 \begin{equation}\label{intensity}
 \langle I_+I_-\rangle =  \int dx C(0)^2 + 2  \int dx_1  dx_2 C\left(\frac{d}{Dc}(s+x_1)\right) C\left(\frac{d}{Dc}(s+x_2)\right)
 = \langle I_+\rangle\langle I_-\rangle + 2 \langle S_+S_-\rangle^2
 \end{equation}
 
Thus, we see that there is a close connection between the amplitude and intensity correlations (for a discussion of this, see the review of Baym~\cite{Baym}).
This explains the fact that the intensity correlation approaches a nonzero value at large $d$ and the collapse of the experimental data also at the intensity level, as seen in Fig.~\ref{fig:ccInt}b.

\begin{figure}
  \begin{center}
    \includegraphics[width=15cm]{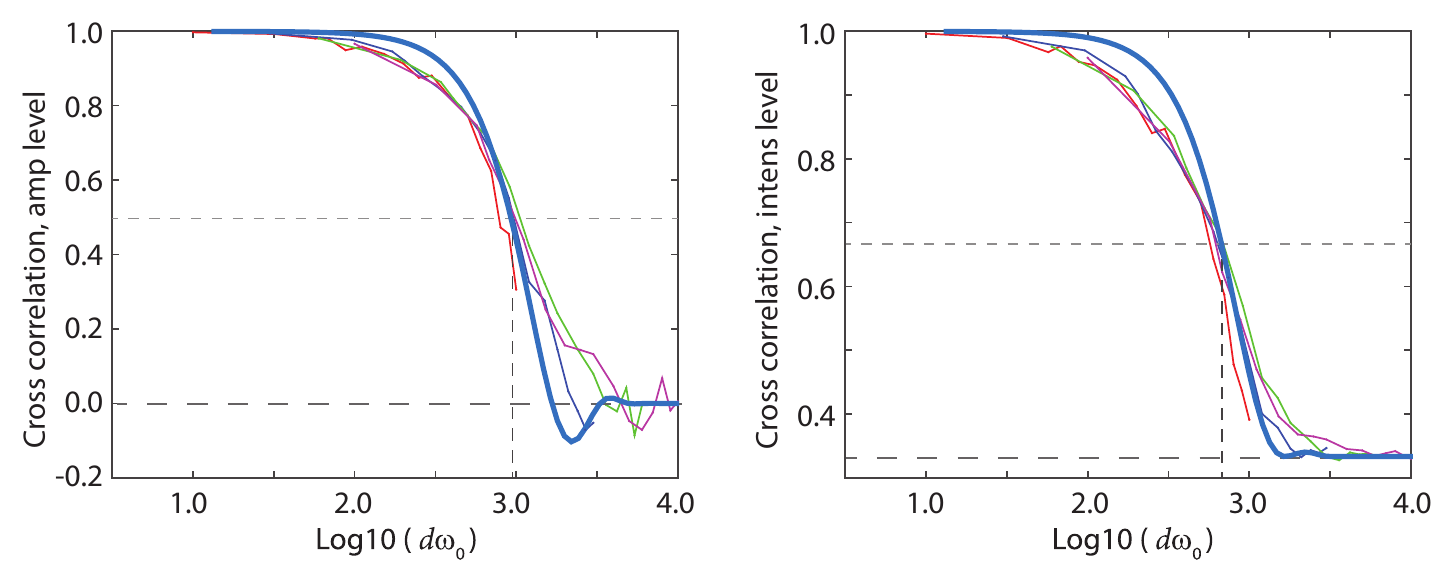}
    \caption{The theoretical predictions (thick lines)
      together with the observed cross correlations. The left figure
      shows the cross correlation of the amplitudes together with the
      prediction from Eq.~\eqref{CCAmp}, while the right figure shows
      the correlation of the intensities. The cross correlation of amplitudes 
      becomes slightly negative before it goes to zero. This weak anti-correlation 
      is easily seen in the theoretical curve, but is also observed experimentally 
      (as well as in Fig.~\ref{fig:CCprinc}). Note that there are no
      fitting parameters, the theoretical curves are based on measured
      parameters.}
      \label{fig:CCwithTheory}
  \end{center}
\end{figure}

It is interesting to compare both the measured cross correlations at the intensity and amplitude level with the theoretical predictions. Since we in the observed correlation functions, Figs.~\ref{fig:ccAmpl}b and \ref{fig:ccInt}b, used the cross correlation normalized to 1 at $d=0$, we rewrite the connection between cross correlations at the amplitude and intensity level, Eq.~\eqref{intensity}, in terms of normalized correlations as
\begin{equation}\label{intensityNorm}
  \langle I_+I_-\rangle  = \frac{1}{3}  + \frac{2}{3} \langle S_+S_-\rangle^2
\end{equation}

In Fig.~\ref{fig:CCwithTheory} we show the theoretical curves, found by numerical integration of \eqref{CCAmp} and using \eqref{intensityNorm}. In the numerical calculations we used the measured width of the waterfall ($L=17$ m), so the curves as presented are without any fitting parameters.  We see that the curves fit well with the observed data, and this means that the observed cross correlation confirms the known width of the waterfall.

If the width was unknown, this would be used as a fitting parameter to be adjusted for optimal fit. We found that this would have occurred for a width of 21 m. This indicates that the method is working as expected while the accuracy is not bad considering the fact that the sound intensity of the sources probably is not uniform across the waterfall.

\section{Summary}

We have through numerical modeling, experimental work on acoustic waves, and with analytic modeling, shown that the cross correlation between the sum-waves hitting two detectors depend on the ratio of path length differences (statistically) and the dominating wavelength of the broadband waves. The results are very similar. Let us take a closer look at the details and compare with the historical HBT values.

The key parameter is the maximum relative time-shift (or maximum relative path length differences) seen for waves originating from the linear source to one of the detectors. It is described by Eq.~\ref{eq:defDelta} along with Figs~\ref{fig:HBTgeometrics} and \ref{fig:CCprinc}.

The value of the parameter $\Delta$ that correspond to a cross correlation of 0.5 (amplitude level) or 0.33 (intensity level), are as follows:

\vspace{6 mm}

\begin{tabular} {l*{6} {c} {r}}
Description           &   Ampl.level  &   Intens.level \\
\hline
Numerical modeling    &     0.30      &       n.a.     \\
Experimental (sound)  &     0.31      &      0.19      \\
Analytical math.      &     0.31      &      0.19      \\
Original HBT, light   &     n.a.      &      0.26      \\
\end{tabular}

\vspace{6 mm}

The value for numerical modeling is based on the calculations leading to Fig.~\ref{fig:CCprinc}b. Both the numerical and analytical results are scalable to waves at all frequencies. The experimental results are based on four different frequencies (within the 100 - 1000 Hz range, $\Delta$ agrees within $\pm \approx$ 10 \%). The 5 degree asymmetry in our experiment has only minor effect on the calculated $\Delta$. 

The $\Delta$ value for the original 1956 HBT work~\cite{HBT} is calculated from an angular extension of Sirius of 0.0068 arc seconds, based on a drop in cross correlation, halfway from max to the limiting value, when the distance between the detectors was 8 meters and a dominating wavelength roughly 500 nm. In this case $\Delta$ is a factor 1.3-1.4 larger than our result. The difference can largely be ascribed the difference in geometry of the source for their broadband waves (circular in the 1956 HBT study and linear in our work).

These results are remarkable since we are working with completely different kinds of waves: Sound waves in air compared to electromagnetic waves in the visual range. The frequencies differ by a factor $10^{12}$, and the wavelengths and apparent angular size of the object as judged from the observer differ by approximately a factor of $10^{7}$.

\section{Discussion}

In principle, our work shows that we may use the HBT effect as a tool to determine the angular extension of a waterfall or other spatially extended sources of independent broadband waves. However, all sound sources will normally be close enough to the observer so that it will be easier to determine the extension of the source by other methods.

The main value of our work is to present a description of the HBT effect that is much easier to understand than both the original classical description as well as the quantum description developed a few years later. Compare for example our description with the description of the HBT effect found in chapter 9 of the textbook ``Waves'' (Berkley physics course)~\cite{Crawford}. Most bachelor-level textbooks do not even present the HBT effect at all.

However, one should not underestimate the difficulties in understanding the difference between mixing of \emph{broadband waves} with mixing of near coherent waves with slightly different frequencies. In our experience, personal numerical modeling experience is a key for a better understanding. It takes time to get acquainted with different descriptions in physics, but it can be worth it's time. Broadband waves are an interesting extension of waves in general. We offer our Matlab programs for everyone. Student projects based on playing around with broadband waves is fully realizable even for second year's bachelor students in physics.

The complexity in existing explanations of the HBT effect are historically conditioned. The experimental effect was based on \emph{intensity} fluctuations at a frequency far from the frequency of the light waves. Thus, intensity fluctuations became the central theme of the theory, and the classical description became rather complex.

Furthermore, in the 1950-ies, it was a strong commitment to describe light as indivisible photons, and the concepts of ``photon bunching'' was not easy to understand. One item in the controversy after the HBT paper on light from Sirius was published, was that the light intensity was so low that it, in mean, could not be more than one photon present in the telescope simultaneously. It has therefore been claimed that ``It is one of the interesting features of [the HBT] result that it cannot be understood in terms of the crude - too crude! - model of a beam of light as a stream of discrete, indivisible, corpuscular photons.''~\cite{Sillito}. Thus, the quantum description is also quite complex.

Our starting point is mixing of broadband waves at the \emph{amplitude} level, at a time resolution fitting to the real frequency of the waves. From this, we can deduce, as a \emph{secondary effect}, the behavior at the intensity level, even at a frequency quite different from the frequency of the wave itself. 

Our model in this paper is a pure classical wave description where ``real waves'' (like sound waves) can be added to each other in all possible proportions. As such, contemplations on our model may influence our, personal ``mental pictures'' of light~\cite{Hentschel}, and lead to renewed personal reflections on aspects like ``Occam's razor'' in the philosophy of science.

It has not escaped our attention that our modelling of the HBT effect may, with proper modifications, provide us with an alternative explanation for one or several of the following phenomena: the Hong, Ou and Mandel effect~\cite{HOM}, the Franson effect~\cite{Franson}, the outcome of the Grangier experiment~\cite{Grangier}, as well as entanglement of photons~\cite{Freedman}, including the various details described for example by Bergli~\cite{Bergli}. Work is in progress.

\bibliographystyle{plain}

\begin{thebibliography}{99}

\bibitem{HBT1954} Hanbury Brown, R. and Twiss, R. Q. (1954) LXXIV. A new type of interferometer for use in radio astronomy, The London, Edinburgh, and Dublin Philosophical Magazine and Journal of Science Series 7, 45, 663-682. Published online: 30 Jul 2010 https://doi.org/10.1080/14786440708520475.

\bibitem{HBT} Hanbury Brown, R., and Twiss, R. Q. 1956. A test of a new type of stellar interferometer on Sirius. Nature 178, 1046-1048.

\bibitem{TLHB1957} Twiss, R.Q., Little, A.G., and Hanbury Brown, R. (1957) Correlation between photons, in coherent beams of light, detected by a coincidence counting technique. Nature 180, 324-326.

\bibitem{Hentschel} Hentschel, K. 2018. Photons. The history and mental models of light quanta. Springer ISBN 978-3-319-95251-2.

\bibitem{Glauber} Glauber, R.J. 2007. Quantum theory of optical coherence. Selected papers and lectures. Wiley-VCH Verlag. ISBN 978-3-527-40687-6. 639 pp.

\bibitem{MandelWolf} Mandel, L., and Wolf, E. 1995. Optical coherence and quantum optics. Cambridge University Press. ISBN 0 521 41711 2. See chapters 9.8 - 9.11 and references therein.

\bibitem{Morten} Caballero, M.D., and Hjorth-Jensen, M. 2018. Integrating a computational perspective in physics courses. arXiv:1802.08871 [physics.ed.ph].

\bibitem{Vistnes2130} Vistnes, A.I. (2014) FYS2130: Oscillations and Waves, Numerical project. Partly described in Vistnes, A.I., (2018) Physics of oscillations and waves, with use of Matlab and Python. Springer, ISBN 978-3-319-72313-6, DOI 19.1007/978-3-319-72314-4, Chapter 15.

\bibitem{Kingsley} Kingsley, A. (2015) Instructional Lab for Undergraduates Utilizing the Hanbury Brown and Twiss Effect. Bachelor project report, Brigham Young University, Dallin Durfee, Advisor.

\bibitem{Zou} Zou, J., Nie, L., Liu, M., and Jiang, C. (2018) Research of Space Positioning Method Based on Sound Field HBT Interference. MATEC Web of Conferences 232, 04028.  https://doi.org/10.1051/matecconf/201823204028  EITCE 2018.

\bibitem{LangForinash} Lang, W.C., Forinash, K. (1998) Time-frequency analysis with the continous wavelet transform. Am. J. Phys. 66, 794-797.

\bibitem{Vistnes} Vistnes, A.I., (2018) Physics of oscillations and waves, with use of Matlab and Python. Springer, ISBN 978-3-319-72313-6, DOI 19.1007/978-3-319-72314-4, Chapter 14.  

\bibitem{Baym} Baym, G. (1998) The physics of Hanbury Brown - Twiss intensity interferometry: from stars to nuclear collisions. arXiv:nucl-th/9804026v2, 24 Apr 1998.

\bibitem{Crawford} Crawford, F.S.Jr. (1968) Waves, Berkley physics course - volume 3. MxGraw-Hill Book Company, New York,

\bibitem{Sillito} Sillito, R.M. (1960) Light waves, radio waves and photons. The institute of Physics Bulletin 11 (5), 129-134. [Note: Sillito was not supporting ``the second quantization'', i.e. that the electromagnetic field itself is quantized, and he argue that we should abolish the word \emph{photon} from or vocabulary.]

\bibitem{HOM} Hong, C.K., Ou, Z.Y., Mandel, L. (1987) Measurement of subpicosecond time intervals between two photons by interference. Phys. Rev. Lett. 59, 2044.  % Mangler sluttsidenr.

\bibitem{Franson} Franson, J.D. (1989) Bell inequality for position and time. Phys. Rev. Lett. 62, 2205.  % Mangler sluttsidenr.

\bibitem{Grangier} Grangier, P., Roger, G., Aspect, A. (1986) Experimental evidence for a photon anticorrelaton effect on a beam splitter: a new light on single-photon interferences. Europhysics Letters 1, 173.  % Mangler sluttsidenr.

\bibitem{Freedman} Freedman, S.J., Clauser, J.F. 1972) Experimental test of local hidden-variable theories. Phys. Rev. Lett. 28, 938-941.

\bibitem{Bergli} Bergli, J., Adenier, G., Th\"{o}rn, A., Vistnes, A.I. (2019) Frequency and phase relations of entangled photons observed by a two-photon interference experiment. Phys. Rev. A 100, 023850.

%\bibitem{Kleppner} Kleppner, D. 2008. Hanbury Brown's steamroller. Physics Today 61. 8.8. https://doi.org/10.1063/1.2970223.

%\bibitem{Bromberg} Bromberg, J.L. 2016. Explaining the laser's light: classical versus quantum electrodynamics in the 1960s. Arch. Hist. Exact Sci. 70, 234-266 (and references therein).

%\bibitem{Shih} Shih, Y. (2011) An introduction to quantum optics. Photon and biphoton physics. Chapter 6. CRC Press, 464 pp.

%\bibitem{blachole} Wielgus, M. et al. (2020) Monitoring the Morphology of M87* in 2009 - 2017 with the Event Horizon Telescope. The Astrophysical Journal, 901:67, https://doi.org/10.3847/1538-4357/. Castelvecchi, D. reported shortly this achievement in Nature 586 (2020) p 18-19.

%\bibitem{Shore} Shore, B.W. (2020) Our changing views of photons. A tutorial memoir. Oxford University Press- ISBN 978-0-19-886285-7.

\end{thebibliography}

\end{document}